\documentclass{aa}
\usepackage{graphicx,txfonts}
\usepackage{natbib}

\newdimen\captwidth
\newdimen\figwidth
\newcommand{\gl}{Gl\,581}

\newcommand{\excs}{\extracolsep{\fill}}
\begin{document}
\title{Dynamical evolution of the Gliese 581 planetary system}
\author{H. Beust \inst{1} \and X. Bonfils \inst{2} \and X. Delfosse
\inst{1} \and S. Udry\inst{3}}
\institute{Laboratoire d'Astrophysique de Grenoble,
UMR 5571 C.N.R.S.,
Universit\'e J. Fourier, 
B.P. 53, F-38041 Grenoble Cedex 9, France  \and
Centro de Astronomia e Astrof\'{\i}sica da Universidade de Lisboa,
Tapada da Ajuda, 1349-018 Lisboa, Portugal \and
Observatoire de Gen\`eve, 51 ch. des Maillettes, CH-1290 Sauverny, Switzerland
}
\date{Received September ..; Accepted December 2, 2007}
\offprints{H. Beust}
\mail{Herve.Beust@obs.ujf-grenoble.fr}
\titlerunning{Dynamical stability of the Gl\,581 system}
\authorrunning{Beust et al.}
\abstract{The M dwarf Gliese 581 has recently been found to harbour
two super Earths
in addition to an already known close-in Neptune-mass planet.
Interestingly, these two planets are considered as potentially
habitable, and recent theoretical works give further credit to this
hypothesis, in particular for the outermost planet (\gl~d).  } {In
this paper, we address the issue of the dynamical stability evolution
of this planetary system. This is important because the basic
stability ensures that a 3-planet model is a physically adequate
description of the radial-velocity (RV) data. It is also crucial
when considering the planets' habitability because the secular
evolution of the orbits may regulate their climate, even in the case
where the system is stable.}
{We have numerically integrated the planetary system
over $10^8\,$yrs, starting from the present fitted solution. We also
performed additional simulations where i) we vary the inclination of the
system relative to the line of sight, ii) assume eccentricities at
the upper limit of the error bars in the radial velocity fit and where iii) we
consider additional (yet undetected) outer planets. We also compute
Lyapunov exponents to quantify the level of dynamical chaos in the
system.}  {In all cases, the system appears dynamically stable, even
in close to pole-on configurations. The system is actually chaotic,
but nevertheless stable. The semi-major axes of the planets are
extremely stable, and their eccentricities undergo small amplitude
variations. 
The addition of potential outer planets does not affect this result.}
{Consequently, from the dynamical point-of-view, a 3-planet model is an
adequate description of the present RV-data set.
Only a limited range of
inclinations can be excluded for coplanar orbits ($i<10^\circ$).  The
climate on the planets is expected to be secularly stable, thus not
precluding the development of life. Gl 581 remains the best candidate
for a planetary system with planets that potentially bear primitive
forms of life.}   \maketitle
\section{Introduction}
\begin{table*}
\caption[]{Orbital parameters of the \gl\ planetary system, as derived
from the fit of \citet{udry07}}
\label{params}
\begin{tabular*}{\textwidth}{@{\excs}lllllll}
\hline\noalign{\smallskip}
Planet & Period (days) & Semi-major axis (AU) & Eccentricity & $\omega$\ (deg)&
$t_\mathrm{p}$ ($\mathrm{JD}-2400000$)& Mass ($M_\oplus$)\\
\noalign{\smallskip}\hline\noalign{\smallskip}
\gl~b & $5.36843\pm0.00031$  & $0.04061\pm0.16\times10^{-6}$ &
 $0.01374\pm0.01405$ & $273.21195\pm60.54198$ & $52998.74631\pm0.90393$ &
$15.82\pm0.25$ \\
\gl~c & $12.92648\pm0.00723$ & $0.07295\pm2.7\times10^{-4}$ &
 $0.15926\pm0.05981$ & $257.41189\pm24.37209$ & $52993.40770\pm1.00130$ &
$5.073\pm0.31$ \\
\gl~d & $83.22730\pm0.65845$ & $0.2525\pm0.013$              &
 $0.12118\pm0.12034$ & $317.01021\pm41.79575$ & $52946.80339\pm11.49888$ & 
$7.804\pm0.69$ \\
\noalign{\smallskip}\hline
\end{tabular*}
\end{table*}
The M dwarf Gliese 581 has been the subject of a recent investigation
with the identification of its 3-planet system. One of the planets
(\gl~b), a Neptune-mass object orbiting the star on a 5.4-day orbit,
has already been known for two years \citep{bonf05}. Recently,
\citet{udry07} have reported the discovery of two additional super
Earths (\gl~c and d), revolving around the star in 12.9 and 83 days
(see details in Table~\ref{params}). Considering Gl581's luminosity,
\citet{udry07} inferred the equilibrium temperature of both planets
and conclude they may lie within the habitable zone of the star.

Detailed further modeling by \citet{bloh07} and \citet{sels07} addresses
the habitability of the planets. Likewise, \citet{bloh07} find
that the greenhouse
effect increases \gl~c temperature so much that they no longer
consider the planet to be in the habitable zone. For similar
reasons, \citet{sels07} find that \gl~c's surface temperature
is very likely higher than the equilibrium temperature calculated by
\citet{udry07}. However, they do not rule out habitability for this
planet, as a large cloud coverage ($>75\%$) would cool down the
planet enough. Conversely, both studies agree that the outermost
planet (\gl~d) is a good candidate for habitability
(although close to the outer edge of the habitable zone)
and actually consider it as the better of the two candidates.

An important and unsettled issue about this system concerns its
dynamical behaviour. It is first important to know whether the
planetary system is dynamically stable and for which range of orbital
inclinations.  If verified, the basic stability of the system ensures
that the model used (3 planets) is a physically adequate description
of the observations (the radial-velocity measurements). If not
verified, the planet detections are not necessarily invalidated (as RV
periodogrammes clearly show that three coherent signals sum up at
specific periods). It would instead mean that either not enough data
were collected to converge toward the ``true'' parameters of the
system or that the model (3 planets) is not complex enough to
describe the data.  Further varying the orbital inclination, we
expect to find that below
a given value -- or, equivalently, above given masses
for the planets -- the system becomes unstable. Not valid physically,
this range of inclinations should be rejected among the possible
solutions. 
This partially constrains the $\sin i$ degeneracy inherent to
radial-velocity detections.

Beyond the basic stability, the secular evolution of the orbits may
play an important role regarding planets' habitability.  All climate
calculations \citep{bloh07,sels07} have been done with the currently
determined orbits.  The secular evolution of the orbits has the
potential of affecting the climate on the planets. A given planet may
lie within the habitable zone but, if subject to significant
eccentricity changes, it can undergo strong climate variations that
eventually preclude life development. The presently determined
eccentricities (Table~\ref{params}) are small enough to ensure climate
stability. But one needs to know which maximum values they reach due
to secular perturbations.

In the present paper, we numerically investigate the secular evolution
of the \gl\ system, starting from the solution of Table~\ref{params}.
In Sect.~2, we study this solution (that we subsequently refer to as the
nominal case). In Sect.~3, we perform other integrations, assuming 
different inclinations from edge-on and letting the initial eccentricities
of the planets reach their maximum values within their error bars.
In order to quantify the dynamical chaos in this system, we
compute Lyapunov exponents for all these solutions in Sect.~4.
In Sect.~5, we investigate the perturbing action of potentially
additional outer planets that have not been detected yet, provided
their contribution to the radial velocity signal is small enough 
compared to the residuals of the 3-planets fit. Our conclusions
are presented in Sect.~6.
\section{The nominal case}
\begin{figure}
\includegraphics[angle=-90,width=\columnwidth]{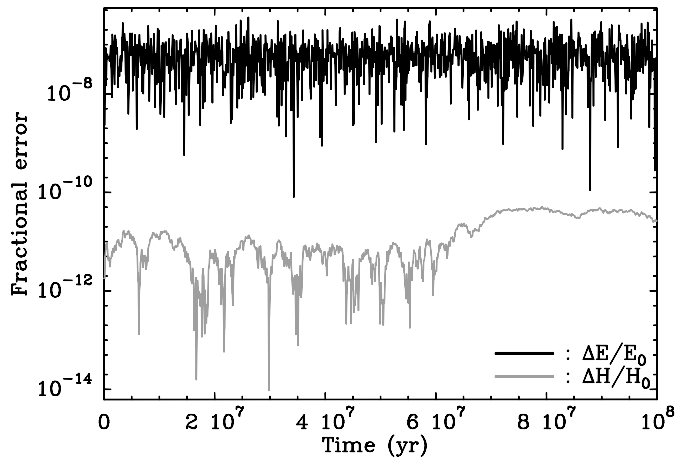}
\caption[]{Fractional errors on the total energy $E$ with respect to
the initial one $E_0$ (black curve), and on the total angular momentum
$H$ with respect to the initial one $H_0$ (grey curve), as a function
of time over the $10^8\,$yr integration} 
\label{energy}
\end{figure}

The best 3-planet orbital fit for \gl\ is explained in
Table~\ref{params}. This solution with the assumptions of coplanarity
and $\sin i = 1$ ($i=90\degr$, an edge-one system) will constitute our
nominal case. We numerically integrate this system taking 
$0.31\,M_\odot$ for the
mass of \gl. The integration is performed using the symplectic 
$N$-body code SyMBA \citep{dun98}, which handles close encounters.
The initial timestep is fixed to
$2\times10^{-4}\,\mbox{yr}=0.18\,\mbox{day}$, i.e. $1/30$ of the
smallest orbital period. Symplectic integration schemes 
usually ensure energy conservation with $10^{-6}$ relative accuracy
as soon as the timestep is taken to $\sim 1/20$ of the smallest orbital
period \citep{ld94}. The integration is carried out over $10^8\,$yr.
On more limited timespans, we checked that taking a significantly
smaller timestep does not change the result. In Fig.~\ref{energy},
we display the fractional errors on the total
energy and angular momentum over the $10^8\,$yr integration. The
energy is preserved to less than $10^{-7}$ relative accuracy. Hence we are
confident in our integration.
\begin{figure*}
\makebox[\textwidth]{
\includegraphics[angle=-90,width=0.32\textwidth]{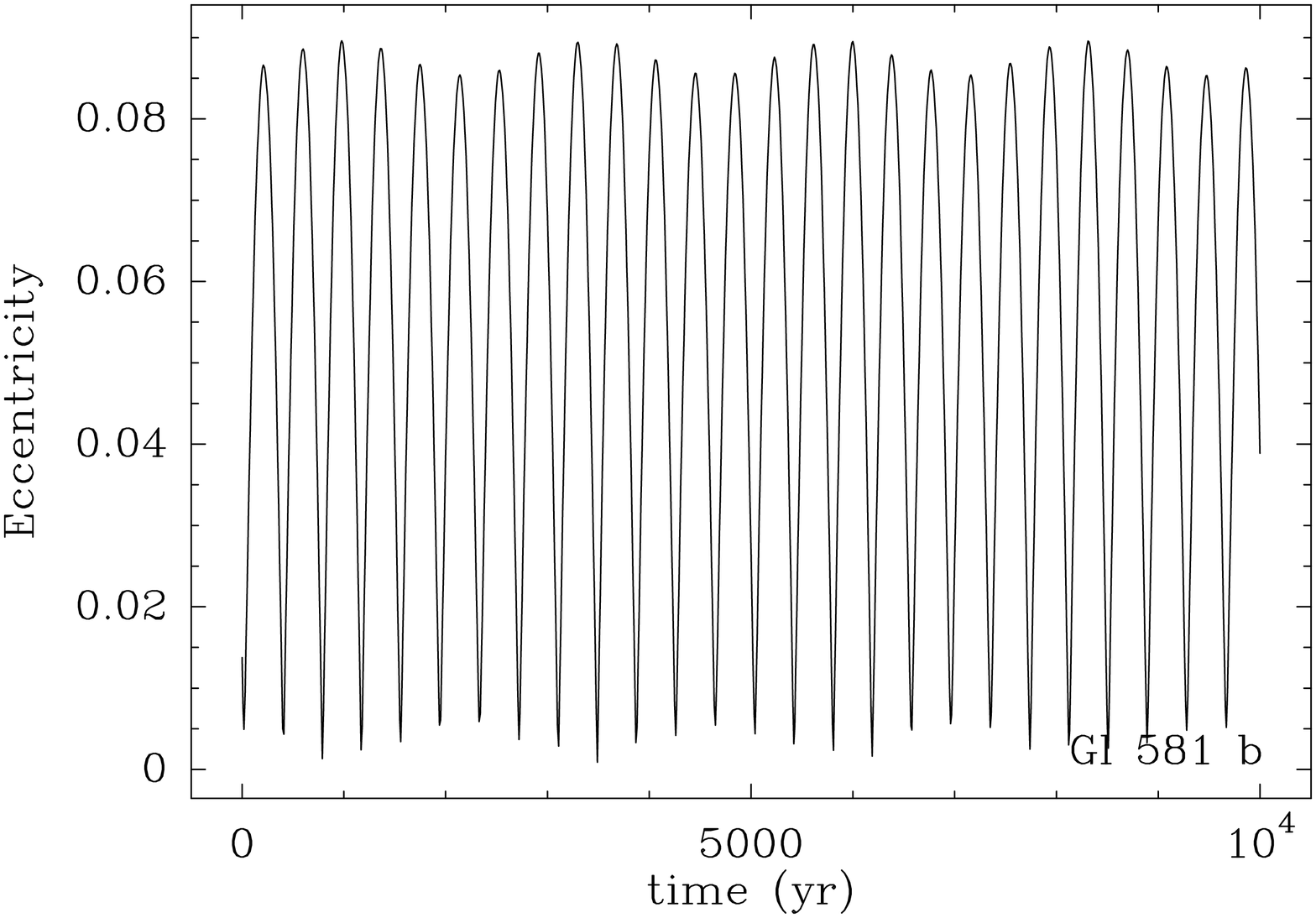} \hfil
\includegraphics[angle=-90,width=0.32\textwidth]{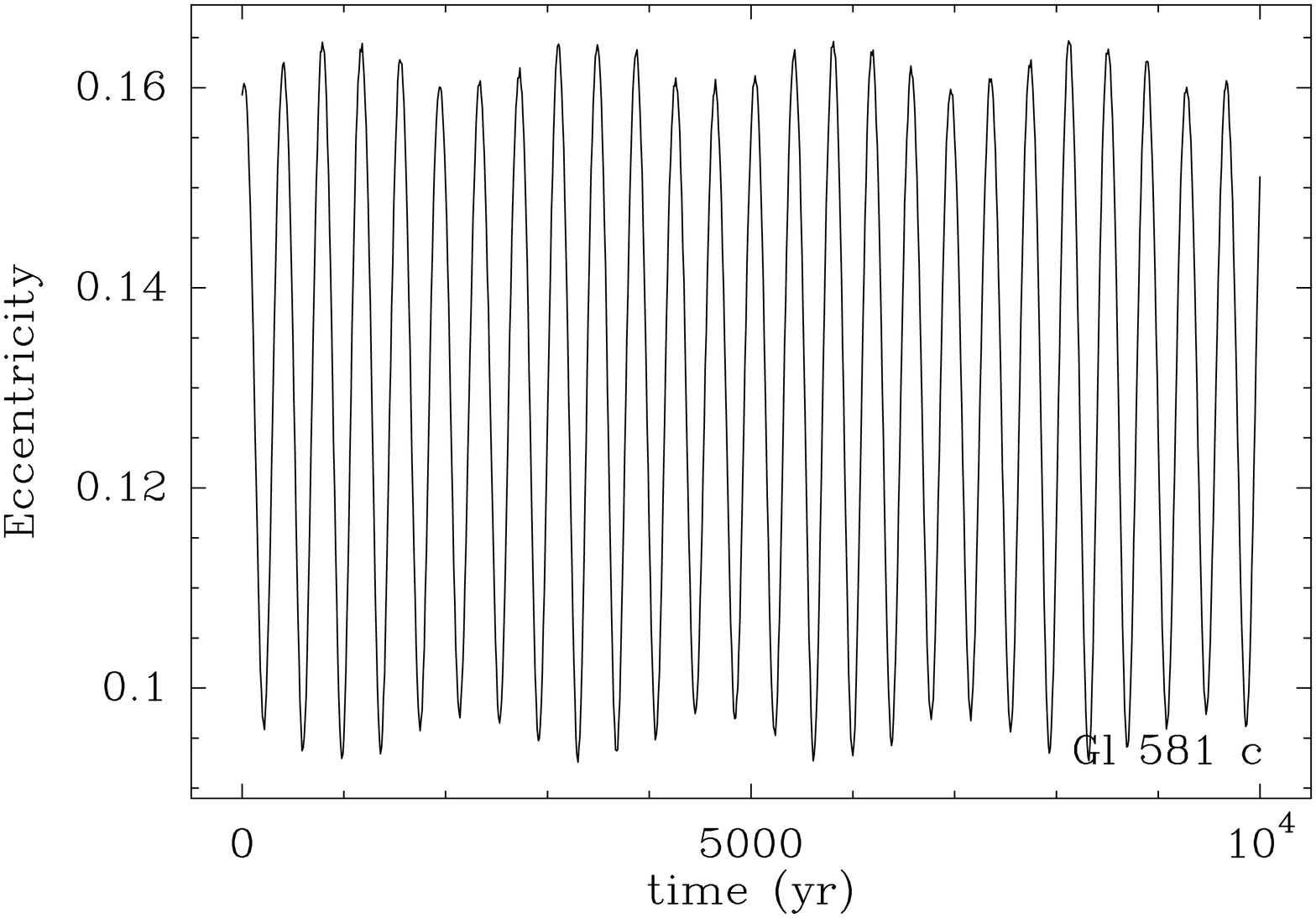} \hfil
\includegraphics[angle=-90,width=0.32\textwidth]{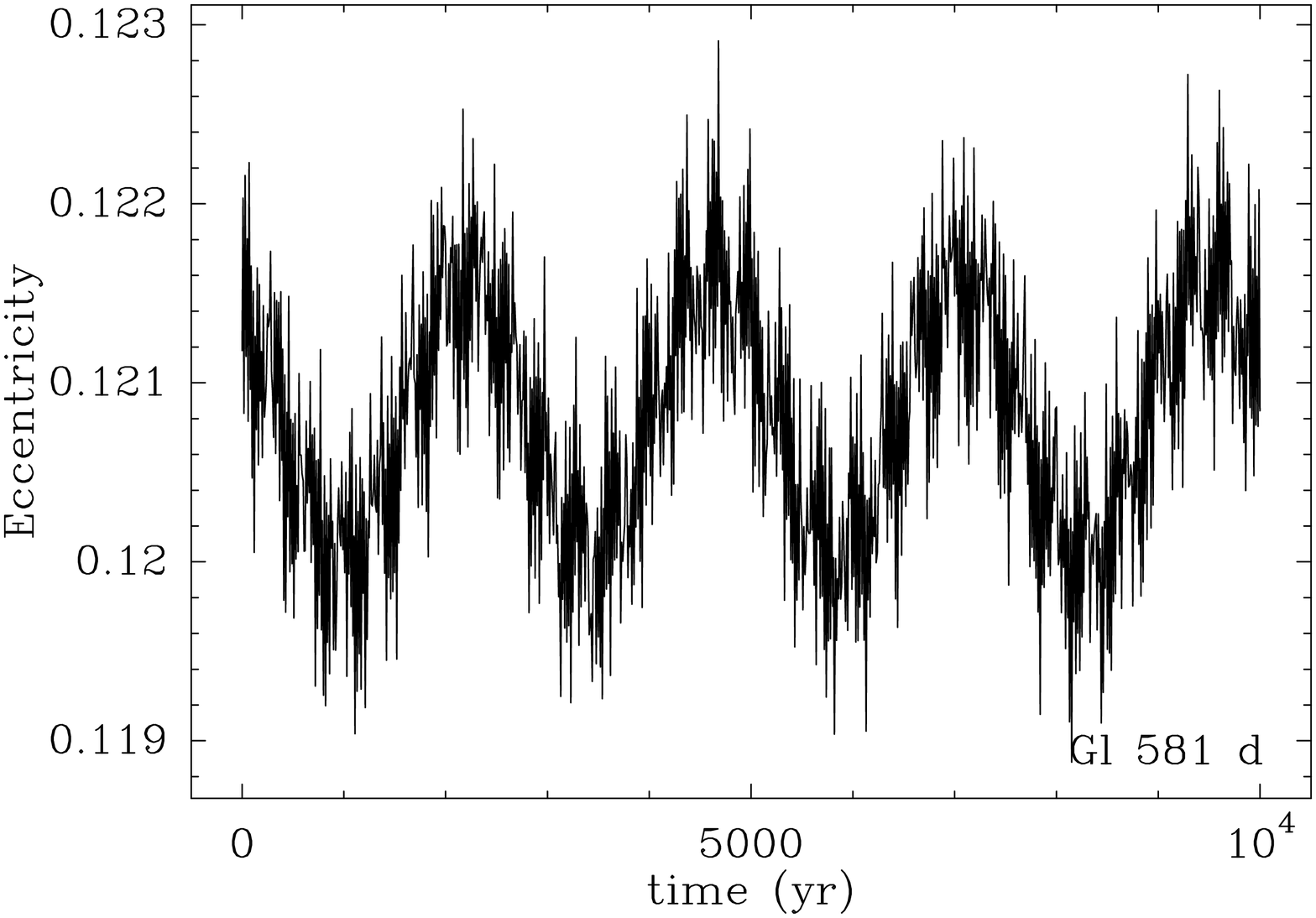}}
\makebox[\textwidth]{
\includegraphics[angle=-90,width=0.32\textwidth]{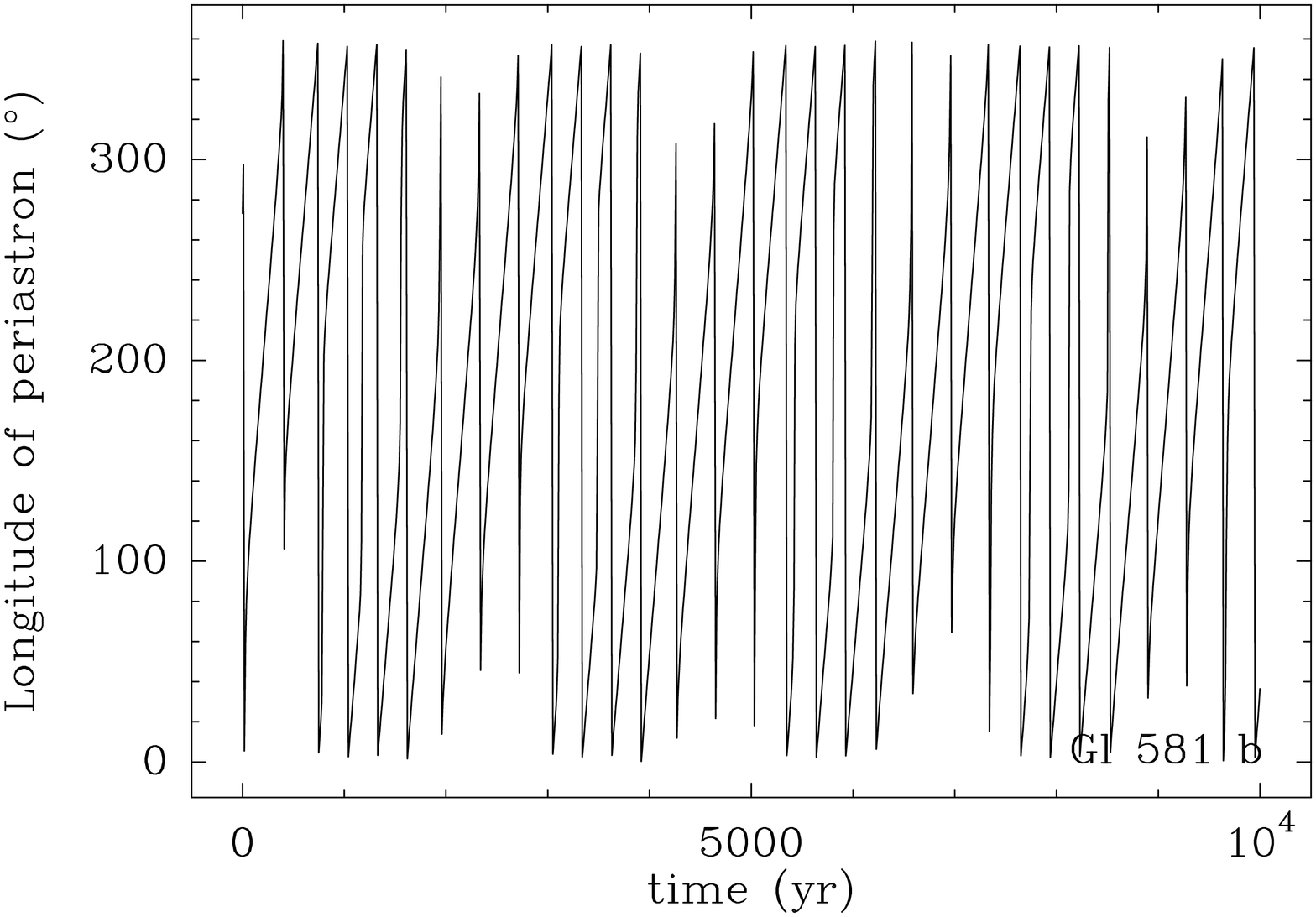} \hfil
\includegraphics[angle=-90,width=0.32\textwidth]{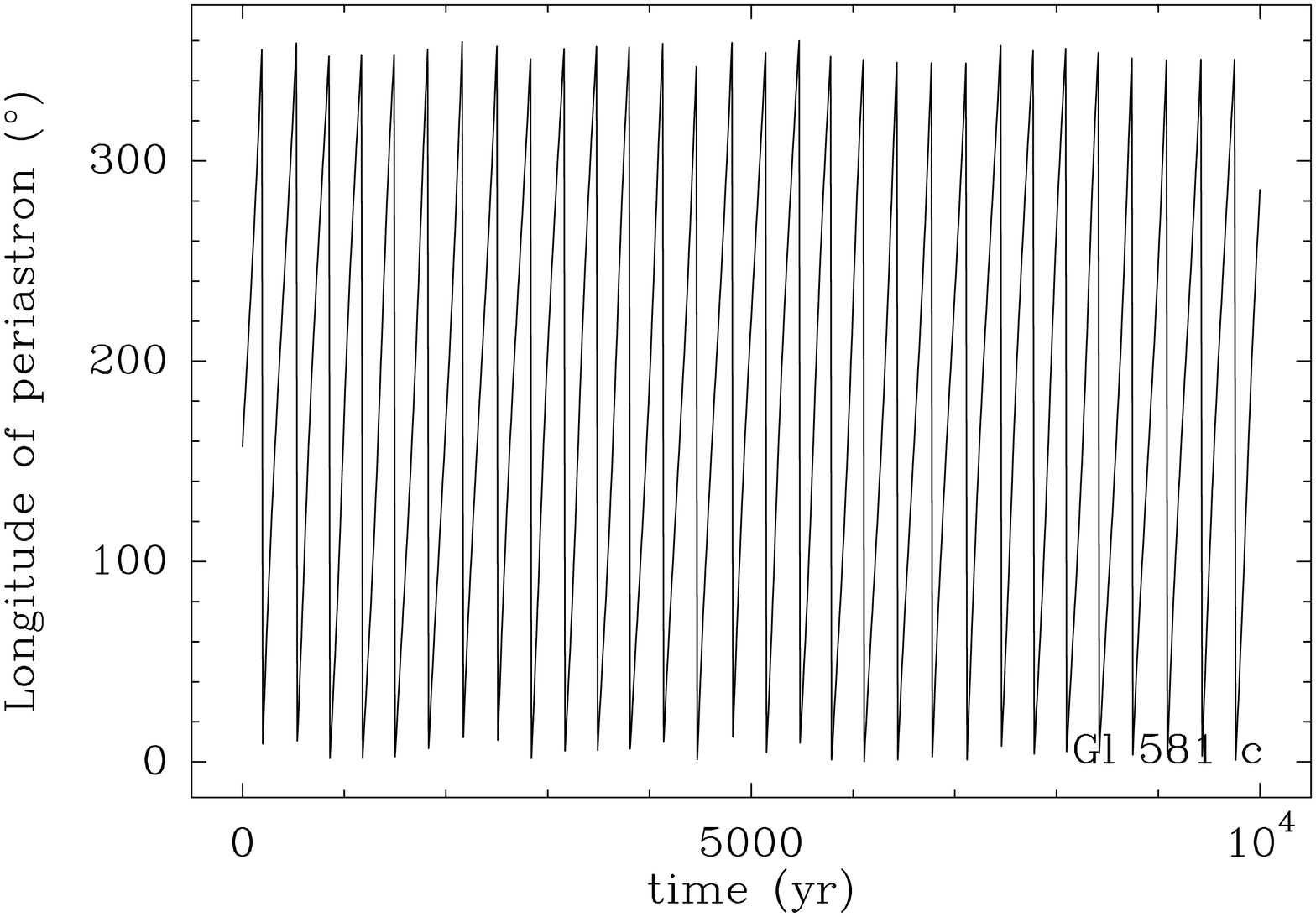} \hfil
\includegraphics[angle=-90,width=0.32\textwidth]{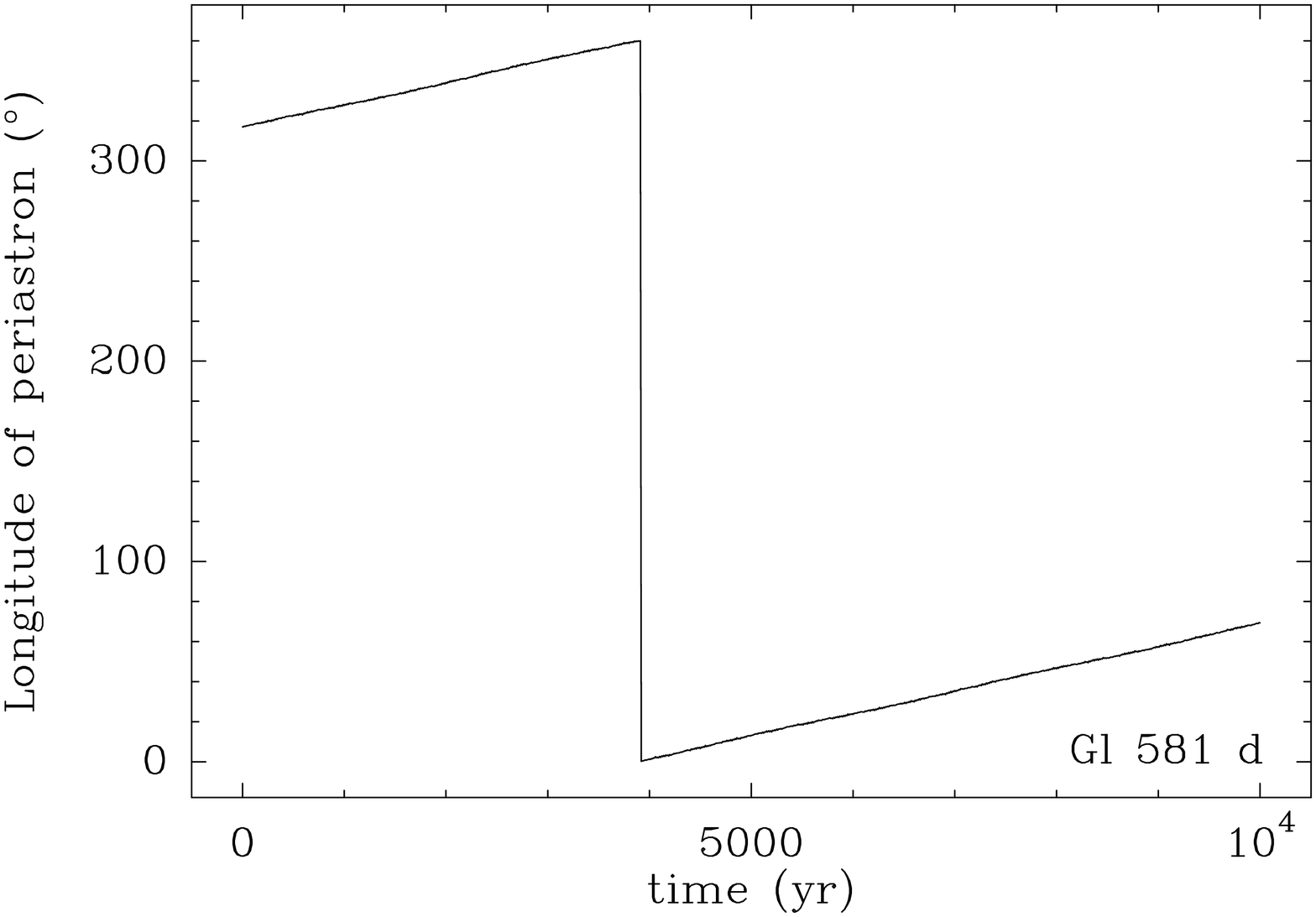}}
\caption[]{The $10^4$ first years of the integration of the nominal
solution. The upper plots show the temporal evolution of the
eccentricity of the three planets \gl~b,c, and d, from left to right,
respectively; the lower plots are the same for the longitude of
periastron $\varpi$}
\label{gl1e4}
\end{figure*}
\begin{table}
\caption[]{Precession frequencies for the nominal solution, as
computed from the linear secular theory}
\label{freqs}
\begin{tabular*}{\columnwidth}{@{\excs}lll}
\hline\noalign{\smallskip}
Name & Frequency (\arcsec/yr)& Period (yr)\\
\noalign{\smallskip}\hline\noalign{\smallskip}
$g_1$ & 3300.9 & 392.62\\
$g_2$ & 539.04 & 2404.2\\
$g_3$ & 38.199 & 33945.\\
\noalign{\smallskip}\hline
\end{tabular*}
\end{table}
\begin{table*}
\caption[]{Variation ranges for some orbital parameters fo the 3
  planets over the $10^8\,$yr integration.}
\label{ranges}
\begin{tabular*}{\textwidth}{@{\excs}lllll}
\hline\noalign{\smallskip}
Planet & Semi-major axis & Eccentricity & Periastron & Apoastron\\
       &    (AU)  & & (AU) & (AU)\\
\noalign{\smallskip}\hline\noalign{\smallskip}
\gl~b & 0.040609 -- 0.046185 & 0.01 -- 0.095 & 0.0368 -- 0.0402 &
0.041 -- 0.0445\\
\gl~c & 0.072885 -- 0.073 & 0.07 -- 0.16 & 0.0614 -- 0.0678 & 0.078 --
00846\\
\gl~d & 0.2522 -- 0.2528 & 0.1200 -- 0.1246 & 0.2207 -- 0.2227 &
0.2823 -- 0.2843 \\
\noalign{\smallskip}\hline
\end{tabular*}
\end{table*}
Figure~\ref{gl1e4} shows the first $10^4$ years of the integration.
We see that the secular variations of the 3 planetary orbits are very
regular. The eccentricities undergo quasi-periodic modulations, while
the longitudes of periastra precess regularly. This solution is
in fact very close to the one we can compute with a linear
secular theory (Laplace -- Lagrange), such as described
the one by \citet{bret74,bret90}. 
In the linear approximation, the secular evolution of the eccentricity
vectors of the 3 planets is a combination of sine and cosine terms
with 3 characteristic frequencies ($g_i$, $i=1,2,3$) that are listed in
Table~\ref{freqs}. These frequencies are obtained by solving the
linear secular equations for their eigenvalues.
We obviously see these characteristic frequencies
in Fig.~\ref{gl1e4}. An interesting outcome is that these precession
frequencies are much higher than in the Solar System, which do not 
exceed $25\arcsec/\mbox{yr}$, and $g_1$ is basically the main precession
frequency of the periastron of \gl~b and \gl~c. This is obviously due
to the much smaller size of the system. Given the error bar on the
fits of the arguments of periastra $(\omega_1, \omega_2)$, the secular
motion should be detectable within $\sim 30\,$years, and probably
less if the orbital fits get more constrained in the near future
thanks to further monitoring. 

In Table~\ref{ranges}, we list the maximum evolution ranges for the
orbital elements of the three planets. The semi-major axes are
extremely stable, revealing a regular dynamics out of any mean-motion
resonance configuration. The evolution ranges of the eccentricities
are narrow, so that we may claim than the system is stable with a high
level of confidence.  While the time span of the integration is
10$^8\,$yr, most of the characteristic features of the secular
evolution of the orbital parameters occur on a 10$^4$\,yr-time
scale. Therefore, even if the star is believed to be older than
$2\times10^9\,$yr, the current integration clearly explores all the
dynamical possible outcomes of the system. Actually, due to the short
orbital periods of the planets (and to the high precession
frequencies), integrating the \gl\ over $10^8\,$yr is basically
equivalent to integrating the Solar System over $\sim 100\,$Gyr !

Interestingly, the present-day
eccentricity of \gl~c roughly corresponds to its maximum values
along its secular evolution, and the eccentricity of \gl~d only has
small variations.
Hence we expect the climate of both outer planets to be secularly stable.
\section{Other solutions}
\begin{figure*}
\makebox[\textwidth]{
\includegraphics[angle=-90,width=0.49\textwidth]{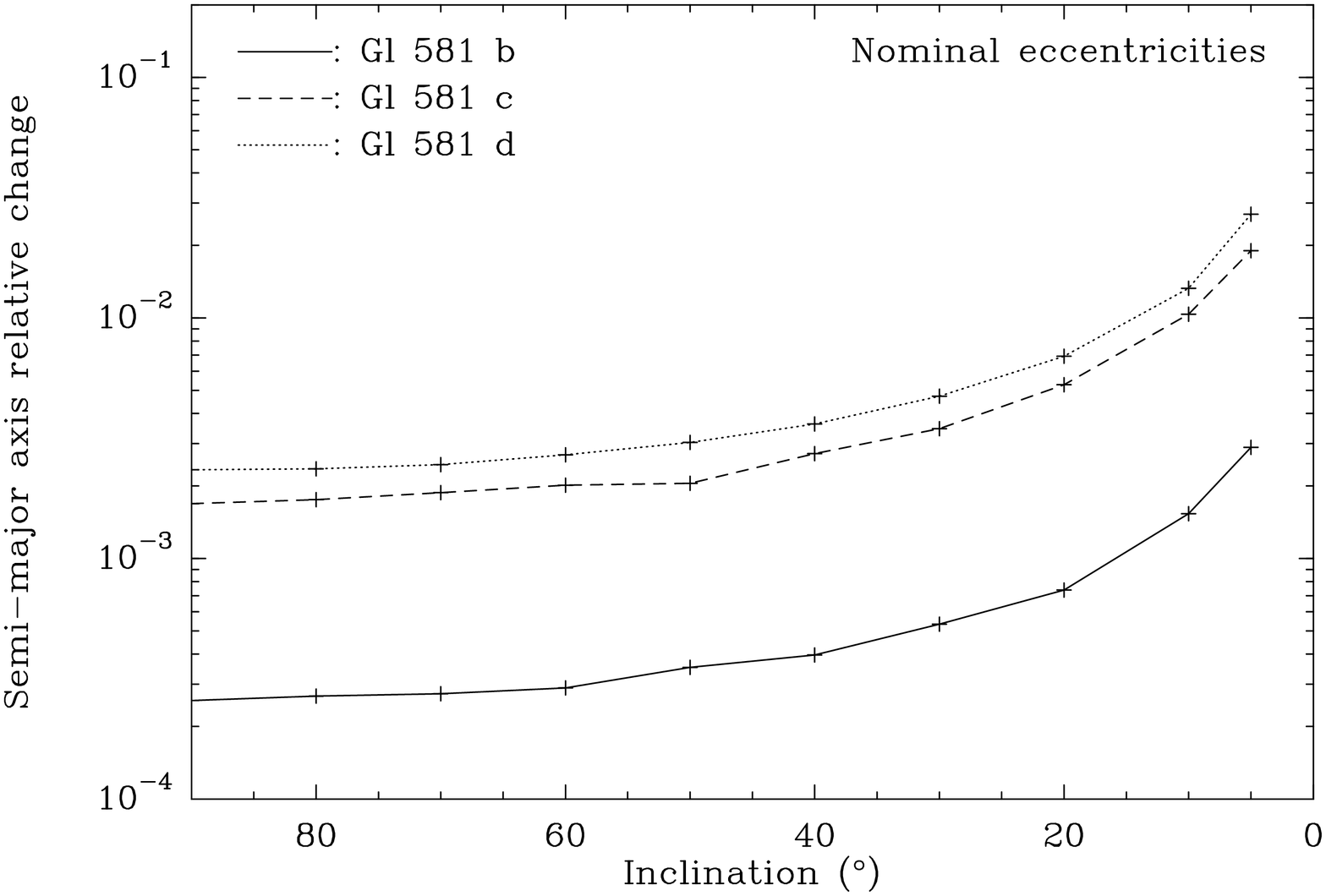} \hfil
\includegraphics[angle=-90,width=0.49\textwidth]{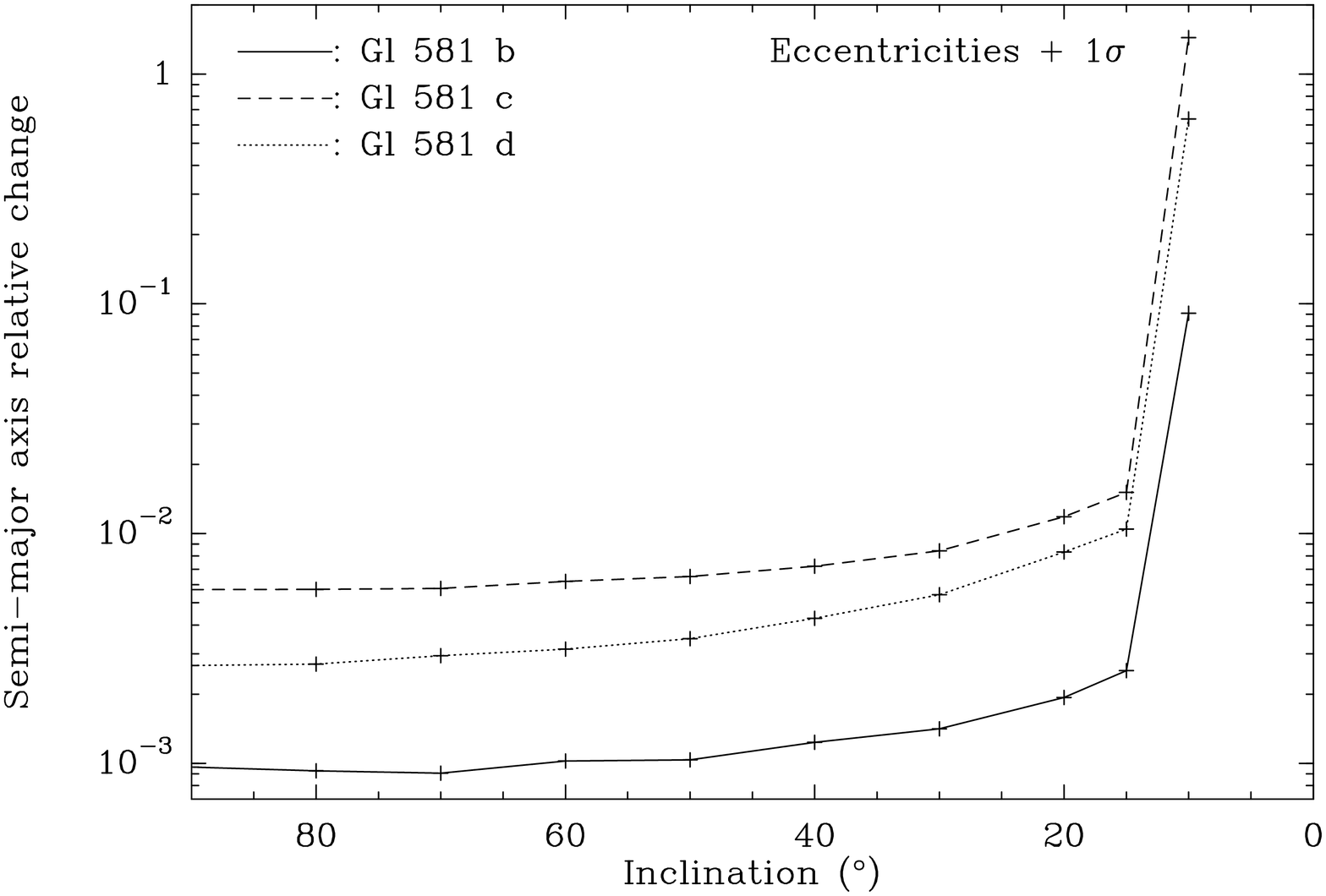}}
\makebox[\textwidth]{
\includegraphics[angle=-90,width=0.49\textwidth]{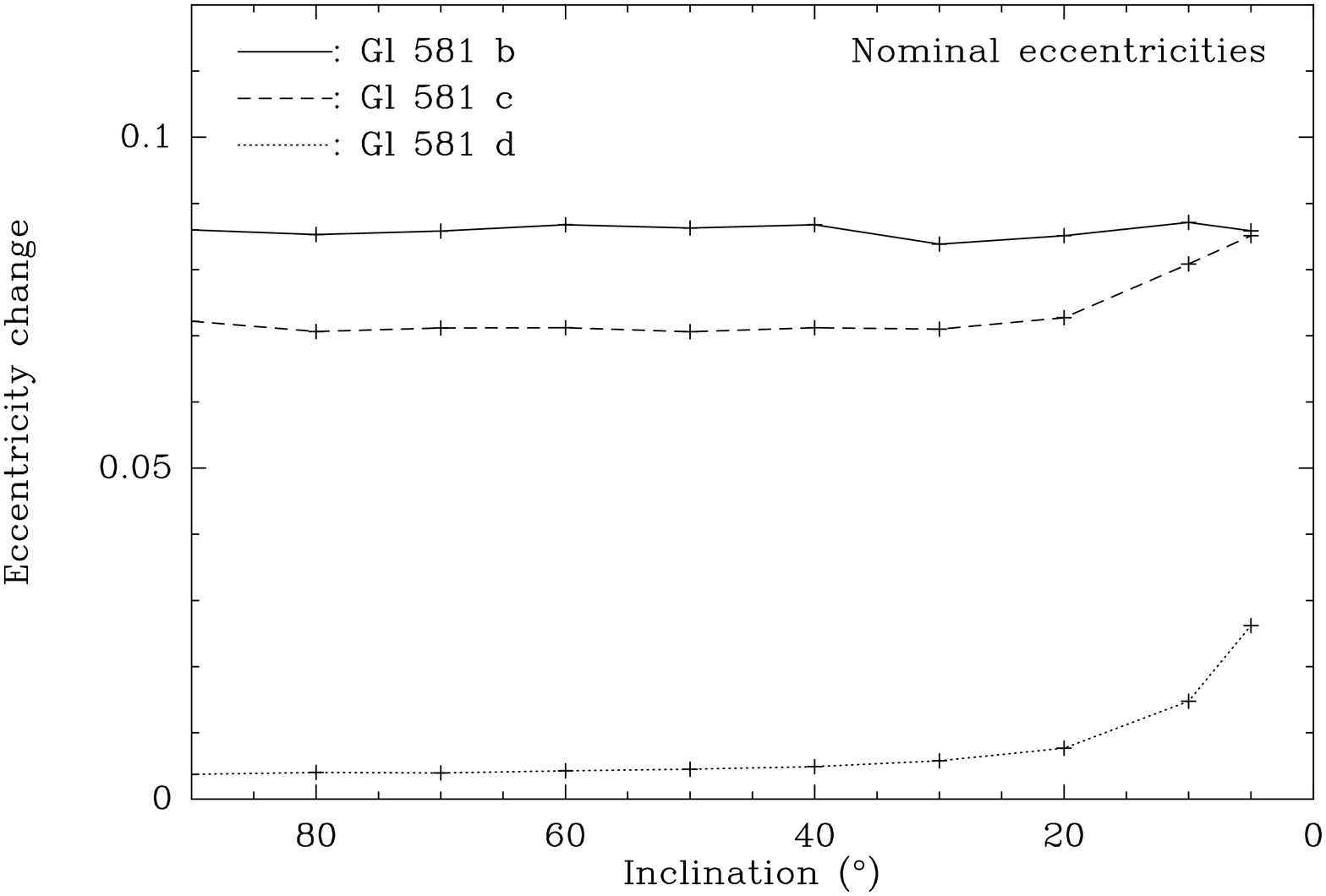} \hfil
\includegraphics[angle=-90,width=0.49\textwidth]{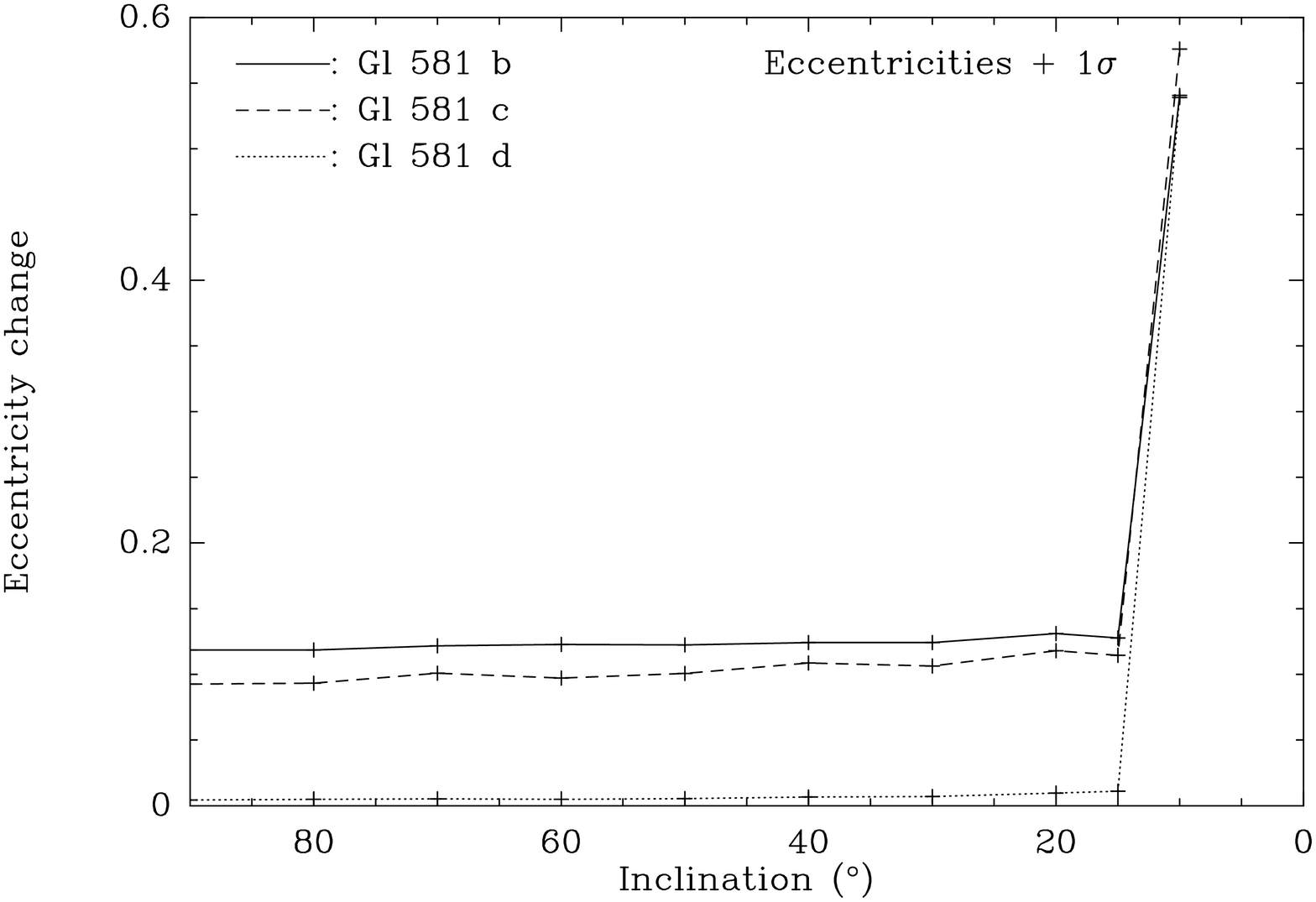}}
\caption[]{Stability of the three-planet system in various
configurations. The maximum variation range for the semi-major
axes (upper plots) and for the eccentricities (lower plots) is displayed
as a function of the assumed viewing inclination of the system with respect
to pole-on. Each cross corresponds to a single simulation. The left
plots correspond to simulations with the nominal eccentricities as
initial conditions, and the right plots to simulations with
eccentricities increased by $1\sigma$ relative to the error bars given
in Table~\ref{params}.}
\label{incli}
\end{figure*}
The nominal solution corresponds to an inclination $i=90\degr$ (so the
lowest possible planetary masses) and to the orbital parameters of
the discovery paper (Table~\ref{params}). Lower inclinations and/or
parameter's values slightly outside the best solution may lead to
different dynamical behaviours that are worth investigating.

In a first set of additional simulations, we assume various
inclinations ranging from 0 (pole on) to $90\degr$ (edge on), but
still holding the initial eccentricities to their nominal values. The
mass of each planet is augmented by a factor $1/\sin i$ with respect
to the nominal case.  In a second set of simulations we assume
different inclinations and, moreover investigate the impact of
eccentricities larger than in the nominal case (as less stability is
only expected if the eccentricity is larger). For that set, we take
the initial eccentricities for the three planets at the upper limit of
their error bars (we add $1\sigma$ to the eccentricities) For both
sets of integrations, we plot the width of the evolution ranges
obtained over the $10^5\,$yr integration for both the semi-major axis
and the eccentricities of the three planets (Fig.~\ref{incli}).

As can be seen from Fig.~\ref{incli}, when the inclination decreases,
the dynamical interactions increase accordingly and
we expect the system to become unstable below a given inclination. As
for the nominal case, the integrations are carried out over
$10^5\,$yr. They naturally show that both the semi-major axis and the
eccentricity take a wider range of values than in the nominal case
with decreasing inclinations. In the first set of integrations
(nominal initial eccentricities), the system nonetheless remains stable
down to $i=10\degr$. Almost pole-on configurations ($i<10\degr$) are
unstable and should be rejected from possible solutions. Although,
such low inclinations are very improbable, and from the statistical
point-of-view, the actual masses of the planets 
are probably close to those listed in Table~\ref{params}.

In the second set of simulations ($1\sigma$ augmented initial
eccentricities), the dynamical
interactions are slightly enhanced and the semi-major axis and the
eccentricity take a wider range of values than for the first set
of additional simulations. The system is therefore found unstable
below larger inclinations ($<20\degr$).

In all cases, the instability appears very unlikely.
If we assume that the rotation axis of the system is randomly 
distributed in space, $i>20\degr$ occurs with a probability of
0.94. In conclusion, irrespective of its actual inclination,
the \gl\ planetary system is very probably stable.
\section{Lyapunov exponents}
\begin{figure}
\includegraphics[angle=-90,width=\columnwidth]{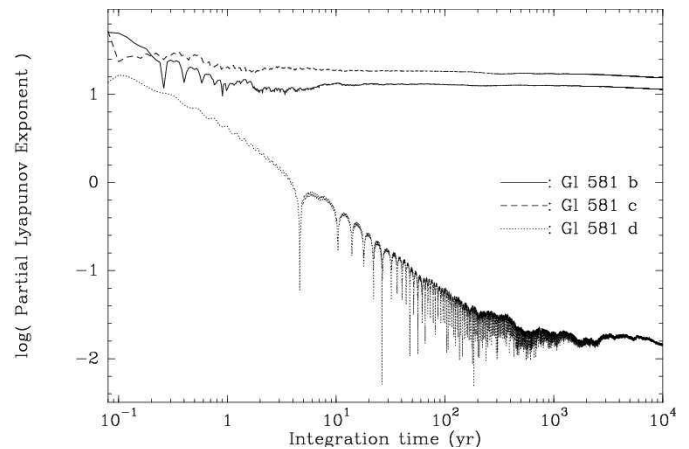}
\caption[]{Progress of the calculation of the partial Lyapunov
exponents as a function of the integration time, in the nominal case
for \gl\ (zero inclination, nominal eccentricities). At $t=10^4\,$yr,
the three exponents have stabilised.}
\label{lyapcalc}
\end{figure}
\begin{figure*}
\makebox[\textwidth]{
\includegraphics[angle=-90,width=0.49\textwidth]{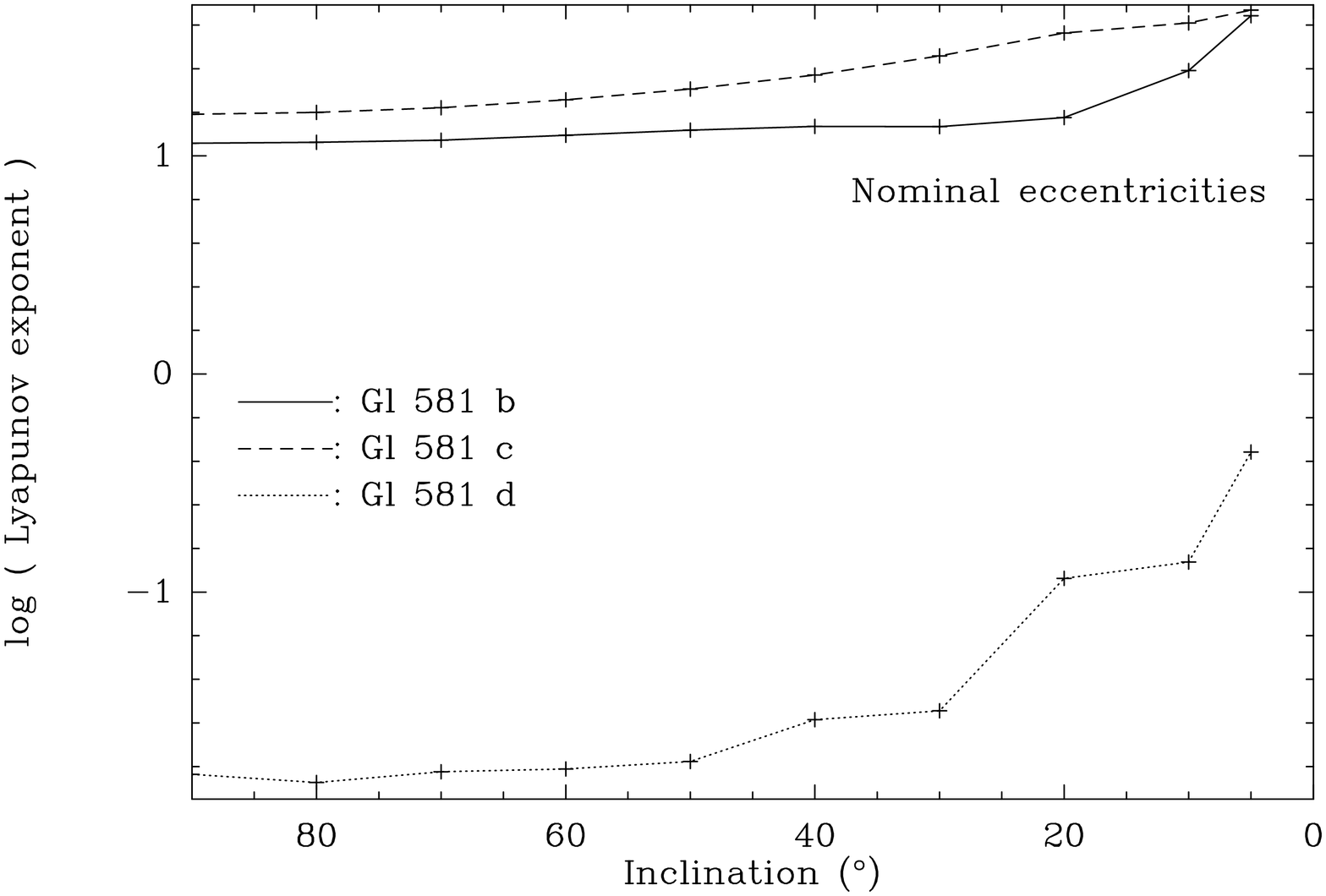} \hfil
\includegraphics[angle=-90,width=0.49\textwidth]{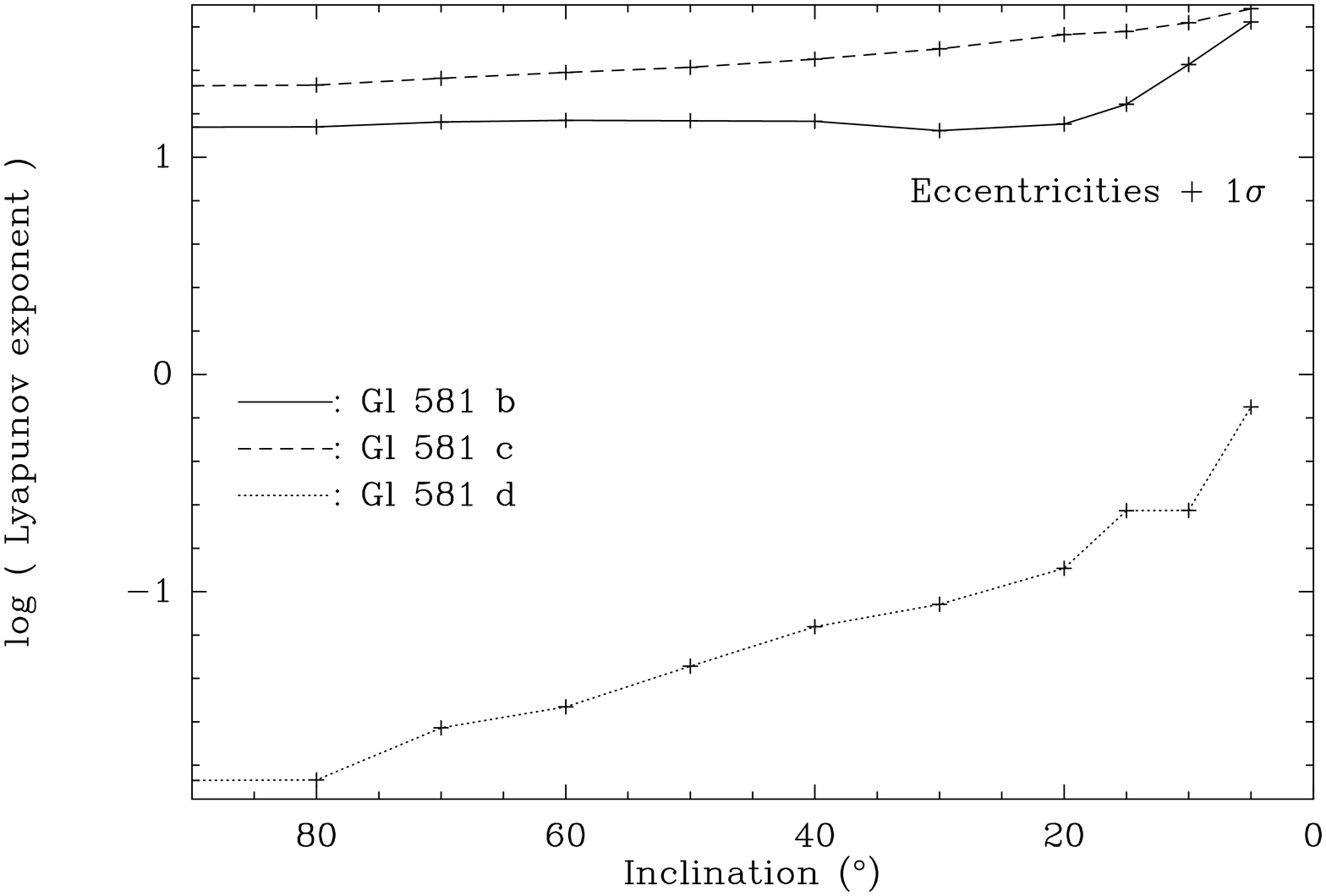}}
\caption[]{Lyapunov exponents as a function of the
inclination $i$, computed for all the simulations described in
Fig.~\ref{incli}. \textbf{Left plot :} simulations with nominal
eccentricities; \textbf{Right plot :} simulations with $1\sigma$
increased eccentricities.}
\label{lyap}
\end{figure*}

Looking for variation ranges for the orbital elements is a basic 
tool for investigating the stability of a planetary system. A more
sophisticated way for quantifying chaos is to compute Lyapunov
exponents. For all simulations decribed above, we compute the
maximum Lyapunov exponents (MLE) for the three planets, following the technique
by \citet{ben80} \citep[see also][]{morby02}. The exponents are
computed by integrating fictitious bodies having initial
conditions that are very close to those of the planets and estimating
their exponential diverge rate. We coupled this algorithm with
the SyMBA integrator. 

When we start integrating the bodies with initial coordinate vector 
$\vec{p_0}$ (holding for the positions and velocities of all the
bodies), we also integrate another system of bodies with identical
masses, but with initial coordinate vector
$\vec{p_0}+\delta\vec{p_0}$, where
$||\delta\vec{p_0}||\ll||\vec{p_0}||$. After a fixed normalization
time $t_\mathrm{norm}$, we compute the error vector
$\delta\vec{p}(t_\mathrm{norm})$
as the difference at $t=t_\mathrm{norm}$ (after integration)
between the coordinate vector of the fictitious bodies and of the 
regular bodies. We then compute
\begin{equation}
s_1=\frac{||\delta\vec{p}(t_\mathrm{norm})||}{||\delta\vec{p_0}||}
\qquad\mathrm{and}\qquad \delta{\vec{p_1}}=
\frac{\delta\vec{p}(t_\mathrm{norm})}{s_1}\quad.
\end{equation}
We then use $\delta\vec{p_1}$ as a new initial error vector for the
fictitious bodies relative to the coordinates vector of the regular
bodies at $t_\mathrm{norm}$, and we iterate the above process.
Each $t_\mathrm{norm}$, the error vector is renormalized this way, and
we obtain a sequence $s_1,s_2,\ldots$ of renormalization factors.
\citet{ben80} proved that the MLE $\mathcal{L}$ can be computed as
\begin{equation}
\mathcal{L}=\lim_{n\rightarrow+\infty}\frac{1}{n\,t_\mathrm{norm}}
\sum_{i=1}^{n}s_i\equiv\lim_{n\rightarrow+\infty}L_n\qquad.
\end{equation}
The result is independent of the choice of $t_\mathrm{norm}$,
provided it is chosen small enough to avoid too large an exponential
divergence. During the integration, we compute $\log L_n$ for every
$t_\mathrm{norm}$, and we try to derive an asymptotic behaviour. Two
cases can occur: i) $\log L_n$ converges towards a finite limit.  Then
the system is chaotic and we have reached $\mathcal{L}>0$ within our
integration time. ii) Or $\log L_n$ keeps decreasing monotonically at the
end of the integration. This means that the orbit is very probably
regular or, more precisely, less chaotic than a given level that
depends on the integration time.

In practice we compute MLEs for all the individual bodies in
the integration: for each orbit we compute the associated (partial)
error vector, and perform individual renormalizations.  Each body has
its own sequence of $s_i$'s. Note that more than the absolute values
of the MLE derived, the comparison between the values derived for the
individual bodies and between different integrations is more relevant.
This shows clearly which orbits are the most chaotic.

In our example applications, we integrate over $10^4\,$yr to compute the
MLEs. $t_\mathrm{norm}$ has been fixed to 0.02\,yr, i.e.,
100 times the timestep. The progress of the computation of the MLEs
in the nominal case for the three planets is plotted as a function of the
integration time in Fig.~\ref{lyapcalc}.
We see that they have all converged towards finite
limits at $t=10^4\,$yr, showing that the orbits are actually chaotic.

The global result of MLE calculation is shown in Fig.~\ref{lyap},
where we have computed the MLEs for the three planets for all the
simulations described in Fig.~\ref{incli} (stopping at
$t=10^4\,$yr). In all cases, we obtain non-zero exponents, showing
that the system is actually chaotic.

We see that the MLEs
slowly increase with decreasing inclinations, showing as expected
that solutions at smaller inclinations are more chaotic, due
to higher planetary masses. We nevertheless note that the
variation is small except for $i<20\degr$. The system is
not much more chaotic at $i=20\degr$ than at $i=90\degr$. This
confirms that there is no real dynamical constraint on the
inclination. We also note that solutions with $1\sigma$ increased
eccentricities are not more chaotic than those with nominal
eccentricities. As a result the dynamical stability does not put
any additional constraint on the planet eccentricities other
than those derived from the radial velocity analysis. 

From Fig.~\ref{lyap}, it also becomes clear that the two inner
planets (\gl~b and c) are much more chaotic than the outer
one (\gl~d) (the exponent is smaller). In fact, the two inner
planets are significantly chaotic. This does not prevent them
from being stable.
Actually, chaos does not necessarily mean instability. The Solar
System is known to be chaotic (but on longer timescales),
yet it is nevertheless stable.
\section{Other planets}
\begin{table}
\caption[]{Semi-major axis and eccentricity variation ranges for the 3
  known planet of \gl, plus additional outer planets (see text), 
  computed over $10^5\,$yr integrations}
\label{addpla}
\begin{tabular*}{\columnwidth}{@{\excs}lll}
\hline\noalign{\smallskip}
Planet & Semi-major axis (AU) & Eccentricity\\
\noalign{\smallskip}\hline\noalign{\smallskip}
\multicolumn{3}{l}{With a $1\,M_\mathrm{J}$ planet at 5\,AU :}\\
\noalign{\smallskip}
\gl~b & 0.04060120 -- 0.04061199 & 0.00367502 -- 0.0895939\\
\gl~c & 0.07286312 -- 0.0729903  & 0.09259277 -- 0.164875\\
\gl~d & 0.25222823 -- 0.25281971  & 0.11884516 -- 0.1228305\\
\noalign{\smallskip}\hline\noalign{\smallskip}
\multicolumn{3}{l}{With a $29.6\,M_\oplus$ planet at 5\,AU :}\\
\noalign{\smallskip}
\gl~b & 0.04060139 -- 0.04061188 & 0.00347993 -- 0.0895106\\
\gl~c & 0.07286522 -- 0.0729892  & 0.09258738 -- 0.1647922\\
\gl~d & 0.25222915 -- 0.2528203  & 0.11887163 -- 0.1226443\\
\noalign{\smallskip}\hline\noalign{\smallskip}
\multicolumn{3}{l}{With a $26.5\,M_\oplus$ planet at 4\,AU :}\\
\noalign{\smallskip}
\gl~b & 0.04060141 -- 0.04061174 & 0.00354749 -- 0.08949917\\
\gl~c & 0.07286447 -- 0.07298892 & 0.09258812 -- 0.16479787\\
\gl~d & 0.25222689 -- 0.25282013 & 0.11884819 -- 0.1226468\\
\noalign{\smallskip}\hline\noalign{\smallskip}
\multicolumn{3}{l}{With a $22.9\,M_\oplus$ planet at 3\,AU :}\\
\noalign{\smallskip}
\gl~b & 0.04060137 -- 0.04061170 & 0.00372438 -- 0.0895144\\
\gl~c & 0.07286369 -- 0.07298806 & 0.09258945 -- 0.16479738\\
\gl~d & 0.25222659 -- 0.25281951 & 0.11884226 -- 0.12267108\\
\noalign{\smallskip}\hline\noalign{\smallskip}
\multicolumn{3}{l}{With a $18.7\,M_\oplus$ planet at 2\,AU :}\\
\noalign{\smallskip}
\gl~b & 0.04060140 -- 0.04061177 & 0.00405747 -- 0.08955927\\
\gl~c & 0.07286632 -- 0.07298852 & 0.09259499 -- 0.16492574\\
\gl~d & 0.25222099 -- 0.25281587 & 0.11886586 -- 0.12266511\\
\noalign{\smallskip}\hline\noalign{\smallskip}
\multicolumn{3}{l}{With a $13.2\,M_\oplus$ planet at 1\,AU :}\\
\noalign{\smallskip}
\gl~b & 0.04060131 -- 0.04061176 & 0.00475611 -- 0.08969945\\
\gl~c & 0.07286289 -- 0.07298820 & 0.09258994 -- 0.16491917\\
\gl~d & 0.25222239 -- 0.25280011 & 0.11882570 -- 0.12246267\\
\noalign{\smallskip}\hline
\end{tabular*}
\end{table}
Our simulations were made with the three known planets orbiting \gl.
However, the system may harbour additional, unknown planets.
The presence of these planets may affect the stability of the
whole system. There are upper limits to the presence of additional
(mainly outer) planets. The maximum amplitude of the residuals
in the 3-planet fits of \citet{udry07} is $\pm 2.1\,\mbox{m\,s}^{-1}$.
Any additional planet should not generate a radial velocity with
a larger amplitude, otherwise it would have already been detected.
Assuming $i=90\degr$ and a circular orbit, this puts severe
constraints on the mass $m$ and distance $d$ of the unseen planet.
We derive
\begin{equation}
\frac{m}{1\,M_\oplus}\leq
13.227\times\sqrt{\frac{d}{1\,\mbox{AU}}}\qquad.
\label{masslim}
\end{equation}
This constraint holds if the unseen planet
generates full-amplitude variations within the timespan of the radial
velocity data, i.e., $\sim 1000\,$days \citep{udry07}. This means that
the orbital period of the unseen planet must not exceed
$\sim$twice this time span to account for this constraint, i.e. an
orbital distance  $d\leq 5.5\,$AU. 
For more distant planets, the constraint is much weaker.
In fact, this upper limit is probably already too large.
Dynamically speaking, we do no expect any hypothetical 
planet orbiting at 5.5\,AU to significantly affect the dynamics of the
inner system located inside 0.25\,AU. We are thus confident in the
conclusions we derive below, as all potentially destabilizing
configurations have been explored.

We thus performed new simulations, each of them with the nominal
conditions, but to which we add an additional planet orbiting the star
on a circular orbit at an arbitrary distance $d$, and with the maximum
mass allowed by Eq.(\ref{masslim}). All the integrations were carried
out over $10^5\,$yr. We did 5 simulations with $d=5$, 4, 3, 2, and
1\,AU. This gives masses of 29.6, 26.5, 22.9, 18.7, and
$13.2\,M_\oplus$, respectively. We also added a simulation with a 1
Jupiter mass ($M_\mathrm{J}$) planet orbiting the star at 5 AU, as the
constraint (\ref{masslim}) is less severe at this distance.  Note that
this case is by far the worst possible disturbing configuration that
is still compatible with the constraints. More distant companions,
even massive, are less destabilizing. In a first-order approximation,
the perturbing effect of a distant planet of mass $m$ orbiting at
distance $d$ on an inner planet orbiting at distance $r$ scales as the
tidal stripping effect on the orbit, i.e. $\propto mr/d^3$. Hence a
$1\,M_\mathrm{J}$ planet at 5\,AU is as disturbing as a
$8\,M_\mathrm{J}$ planet at 10\,AU and a $32\,M_\mathrm{J}$ brown
dwarf at 20\,AU. The AO surveys would likely have already detected
such a massive companion.

In all cases, the whole system appears just as stable as without any 
additional planet. The result concerning the stability is summarised
in Table~\ref{addpla} where we give the semi-major axis and
eccentricity variation ranges for the three known planets. 
The results are very similar among the 6 different integrations,
even in the case of a Jovian planet,
showing that the additional planet has little influence on the
stability of the inner system. Moreover, Table~\ref{addpla} is easily
compared to Table~\ref{ranges}. The varation ranges are very similar.
Therefore, we may stress that any additional outer planet that
fits into the constraint of the radial velocity residuals
does not affect the stability of the 3-planet system.
Note that the maximum eccentricity values in
Table~\ref{addpla} are actually slightly lower than those in
Table~\ref{ranges}. This could appear surprising, since the integration
in Table~\ref{ranges} is made without any additional perturber. 
Recall, however, that it extends over $10^8\,$yr instead of 
$10^5$ for those in Table~\ref{addpla}. This shows conversely that if we were
entending the integrations of Table~\ref{addpla} up to $10^8\,$yr, we
should expect slightly wider variation ranges. The basic conclusion 
nevertheless remains: the stability of the system is not affected.

\section{Discussion}
We have computed the secular evolution of the \gl\ planetary system in
various possible configurations. The main conclusion is that the
system is almost always stable. It is stable for inclinations as low
as $\sim 20\degr$ and even if the initial eccentricities are augmented by
their 1-$\sigma$ error bars.

As expected for any planetary system with regular dynamics, the
semi-major axes vary very little and the three planets are expected to
remain at their current location with respect to the star. Meanwhile,
the eccentricities of the two outer planets (both considered for
habitability) reach values that are significantly above the Earth's
value.  Concerning \gl~c, the present-day eccentricity is close to its
maximum value. This planet is not expected to get much farther away
from its parent star and, to maintain a surface temperature cool
enough to allow the presence of liquid water, a high water-cloud
coverage ($\sim$75\%) would be required at any time.  Regarding \gl~d,
the nominal eccentricity is non negligible ($\sim 0.12$) and also
found to be very stable. It is significantly above the maximum value
reached by the Earth throughout its secular evolution
\citep[$\sim0.06$, see e.g.][]{lask88} and corresponds to a 24\%
variation of the radiation flux received from the star between
apoastron and periastron. The anomalistic season effects should
therefore be strong, if not damped by the short orbital period (83
days). If we compare the periastron and apoastron values of \gl~d to
the habitable zone calculations by \citet{sels07} and \citet{bloh07},
we see that \gl~d is outside the habitable zone at apoastron but well
inside at periastron. As pointed out by \citet{sels07}, the average
stellar flux received by an eccentric orbit is enhanced by a factor
$1/\sqrt{1-e^2}$ with respect to a circular orbit with the same
semi-major axis. This can help maintaining \gl~d in the habitable
zone. What we show here is that this effect is secularly permanent.

Now, if the obliquity of the rotation axis of this planet is non-zero,
this should combine with the obliquity's seasonal effect and lead to
climate differences between the hemispheres of this planet, much like
Mars presently. The obliquity of \gl~d is of course unknown, but
\citet{sels07} and \citet{bloh07} agree in claiming that, given the
estimated age of the star ($>$2~Gyrs), the rotation of \gl~d should
already be tidally locked with the orbital motion.  In that case, we
would expect the obliquity to have been set to zero by tidal effects,
and there should instead be climate differences between the night and
day hemispheres.  This could help in maintaining the day hemisphere
habitable.  \citet{sels07} also show that tidal locking does not
contradict the non-zero eccentricity of the orbit. Tides usually tend
to both synchronize the rotation and circularize the orbit. The
circularization time is almost always longer than the synchronization
time \citep{hut81}.  For \gl~d, \citet{sels07} estimate the
synchronization time to 10\,Myrs and the circularization time to
10\,Gyrs, i.e. well above the present age of the system.
\begin{acknowledgements}
All the computations presented in this paper were performed at the 
Service Commun de Calcul Intensif de l'Observatoire de Grenoble (SCCI).
\end{acknowledgements}

\end{document}